\documentclass[aps,prd,twocolumn,amssymb,eqsecnum,showpacs,showkeyes,secnumarabic,graphics,floatfix,nofootinbib,tightenlines,longbibliography,superscriptaddress]{revtex4-1}

\usepackage{graphicx}
\usepackage{epsfig} 
\usepackage{bm}
\usepackage{cancel}  
\usepackage{dcolumn}  
\usepackage{amsmath} 
\usepackage{hhline} 
\usepackage[dvipsnames]{xcolor}
\usepackage[utf8]{inputenc}
\usepackage{cancel}
\usepackage[dvipsnames]{xcolor}  
\usepackage{float}
\usepackage{longtable}
\usepackage[breaklinks=true,colorlinks=true,
linkcolor=blue,urlcolor=blue,citecolor=MidnightBlue,
bookmarks=true,bookmarksopenlevel=2]{hyperref}
\usepackage{mathrsfs}

\def \beq  {\begin{equation}}
\def \eeq  {\end{equation}}
\def \ber  {\begin{eqnarray}}
\def \eer  {\end{eqnarray}}

\begin{document}
\newcommand{\newc}{\newcommand}

\newc{\be}{\begin{equation}}
\newc{\ee}{\end{equation}}
\newc{\ba}{\begin{eqnarray}} 
\newc{\ea}{\end{eqnarray}}
\newc{\bea}{\begin{eqnarray*}}
\newc{\eea}{\end{eqnarray*}}
\newc{\D}{\partial}
\newc{\ie}{{\it i.e.} }
\newc{\eg}{{\it e.g.} }
\newc{\etc}{{\it etc.} }
\newc{\etal}{{\it et al.}}
\newc{\lcdm}{$\Lambda$CDM }
\newc{\lcdmnospace}{$\Lambda$CDM}
\newcommand{\nn}{\nonumber}
\newc{\ra}{\Rightarrow}
\newc{\omm}{$\Omega_{m}$ }
\newc{\ommnospace}{$\Omega_{m}$}
\newc{\fs}{$f\sigma_8$ }
\newc{\fsz}{$f\sigma_8(z)$ }
\newc{\fsnospace}{$f\sigma_8(z)$}
\newc{\plcdm}{Planck/$\Lambda$CDM }
\newc{\plcdmnospace}{Planck15/$\Lambda$CDM}
\newc{\wlcdm}{WMAP7/$\Lambda$CDM }
\newc{\wlcdmnospace}{WMAP7/$\Lambda$CDM}
\newcommand{\fss}{{\rm{\it f\sigma}}_8}

\title{Weak gravity on a $\Lambda$CDM background}
\author{Radouane Gannouji}\email{radouane.gannouji@pucv.cl}
\affiliation{Instituto de F\'{\i}sica, Pontificia Universidad Cat\'olica de Valpara\'{\i}so,
Av. Brasil 2950, Valpara\'{\i}so, Chile}
\author{Leandros Perivolaropoulos}\email{leandros@uoi.gr}
\affiliation{Department of Physics, University of Ioannina, GR-45110, Ioannina, Greece}
\author{David Polarski}\email{david.polarski@umontpellier.fr}
\affiliation{Laboratoire Charles Coulomb, Universit\'e de Montpellier \& CNRS UMR 5221, F-34095 Montpellier, France}
\author{Foteini Skara}\email{f.skara@uoi.gr}
\affiliation{Department of Physics, University of Ioannina, GR-45110, Ioannina, Greece}

\date {\today}  

\begin{abstract}
We consider Horndeski modified gravity models 
obeying stability, velocity of gravitational waves $c_T$ equals $c$ and quasistatic approximation (QSA) on subhorizon scales. 
We assume further a $\Lambda$CDM background expansion and a monotonic evolution on the cosmic background of the $\alpha$ functions as $\alpha_i= \alpha_{i0}~a^s$ where $i=M,B$, $a$ is the scale factor and $\alpha_{i0}$ ($\alpha_{M0}, \alpha_{B0}$), $s$ are arbitrary parameters. 
We show that the growth and lensing reduced (dimensionless) gravitational couplings $\mu\equiv G_{\rm growth}/G$, $\Sigma\equiv G_{\rm lensing}/G$ exhibit the following generic properties today:  $\Sigma_0 < 1$ for all viable parameters, $\mu_0<1$ (weak gravity today) is favored for small $s$ while $\mu_0>1$ is favored for large $s$. 
We establish also the relation $\mu\geq \Sigma$ at all times.
Taking into account the $f\sigma_8$ and $E_G$ data constrains the parameter $s$ to satisfy $s\lesssim 2$. Hence these data select essentially the weak gravity regime today ($\mu_0<1$) when $s<2$, while $\mu_0>1$ subsists only marginally for $s\approx 2$. 
At least the interval $0.5\lesssim s \lesssim 2$ would be ruled out in the absence of screening. 
We consider further the growth index $\gamma(z)$ and identify the $(\alpha_{M0},\alpha_{B0},s)$ parameter region that corresponds to specific signs of the differences $\gamma_0-\gamma_0^{\Lambda CDM}$, and $\gamma_1-\gamma_1^{\Lambda CDM}$, where $\gamma_0\equiv \gamma\bigl|_{z=0}$ and $\gamma_1\equiv \frac{{\rm d}\gamma}{\rm d z}\bigl|_{z=0}$. 
In this way important information is gained on the past evolution of $\mu$. 
We obtain in particular the signature $\gamma_0>\gamma_0^{\Lambda CDM}$ for $s<2$ in the selected weak gravity region. 
\end{abstract}
\maketitle 

\section{Introduction}
\label{sec:Introduction}
The standard Lambda Cold Dark Matter ($\Lambda$CDM) model 
is a simple and generic model that has been shown to be consistent with a wide range of cosmological observations including geometric and dynamical probes. 
Despite its successes, $\Lambda$CDM is confronted with challenges at both the theoretical and the observational level. At the observational level, it faces in particular the following tensions: $H_0$ tension, growth tension, Cosmic  Microwave  Background (CMB) high-low $l$ tension \cite{Addison:2015wyg,Aghanim:2019ame}, Baryon  Acoustic  Oscillation (BAO) ly-$\alpha$ tension \cite{Addison:2017fdm}, suppressed high angular scale correlation in CMB temperature maps \cite{Bernui:2018wef}, hints for violation of statistical isotropy in the CMB maps 
\cite{Schwarz:2015cma}, for a comprehensive list of difficulties on small scales see e.g. \cite{Bullock:2017xww} . The $H_0$ tension is based on the fact that the CMB measured value of the Hubble parameter $H_0$ \cite{Ade:2015xua,Aghanim:2018eyx} assuming $\Lambda$CDM is significantly lower (about $4\sigma$) than the value indicated by local distance ladder measurements from supernovae \cite{Riess:2016jrr,Riess:2018byc} and lensing time delay indicators \cite{Birrer:2018vtm}, with local measurements suggesting a higher value. The growth tension is based on the fact that the observed growth of cosmological perturbations is weaker than the growth predicted by the standard \plcdm  parameter values. 
These tensions, if not due to statistical or systematic errors, may indicate the need for additional degrees of freedom extending $\Lambda$CDM.
A generic physically motivated origin of such degrees of freedom is the extension of General Relativity (GR) to Modified Gravity (MG) models.  
Actually consideration of MG models is not restricted to solving these tensions and has been introduced to produce the late-time accelerated expansion itself. 
A great variety of MG models have been proposed so far to account for these tensions, in particular for the $H_0$ 
and the growth tensions.  
A wide class of such MG theories is provided by Horndeski gravity. Horndeski gravity models  
\cite{Horndeski:1974wa,Deffayet:2011gz}
(see e.g. \cite{Kase:2018aps,Kobayashi:2019hrl} for a comprehensive review) is the most general Scalar-Tensor (ST) theory involving a scalar degree of freedom in four dimensions with  second order equations of motion therefore avoiding the Ostrogradsky instability
\cite{Ostrogradsky:1850fid,Woodard:2015zca}.
It provides a general framework to construct models of dark energy as well as inflation. 
It includes dark energy models inside GR such as quintessence 
as well as a wide variety of modified gravity models, such as $f(R)$ gravity \cite{DeFelice:2010aj},
Brans-Dicke (BD) theories \cite{Brans:1961sx,DeFelice:2010jn}, galileons etc. 
However, the recent detection of gravitational waves emitted by binary systems
has imposed stringent constraints on their speed $c_T$ constraining the latter to be extremely close to the speed of light $c$
($c_T/c=1\pm 10^{-15}$) 
\cite{TheLIGOScientific:2017qsa,Goldstein:2017mmi}.
Remember that $c_T=c$ is a fundamental prediction of GR. This constraint has significantly restricted the observationally allowed subclasses of Horndeski models. Notice that a way to get around this constraint is to assume ab initio that $c_T$ depends on its wavelength \cite{deRham:2018red}. 
The gravitational properties of Horndeski theories can be elegantly expressed by means of four free independent functions of time namely the $\alpha$-basis $\alpha_i(t)$ $(i=M,K,B,T)$, (see Ref.\cite{Bellini:2014fua}), describing the linear perturbations, while the background expansion is given by the Hubble parameter $H(a)$ where $a$ is the scale factor. These four time dependent phenomenological functions describe any departure from GR and also  characterize specific physical properties  of the  Horndeski models. GR is recovered when all $\alpha_i$ are set to zero.

For specific choices of the $\alpha_i$ the resulting theory may be unstable on a given background $H(a)$. Thus two types of instabilities may occur:
\begin{itemize}
\item
Ghost instabilities \cite{Sbisa:2014pzo} which arise when the kinetic term of the background perturbations has the wrong sign 
giving negative energy modes. In this case the high energy vacuum is unstable with respect to the spontaneous production of particles. 
\item
Gradient instabilities 
which arise when the background $H(a)$ evolves in a region where the sound speed of the perturbations becomes imaginary ($c_s^2<0$). This leads to the appearance of exponentially growing modes of the form $e^{c_skt}$ at  small scales. 
\end{itemize}
The functions $\alpha_i$ of a physically acceptable Horndeski model should avoid such instabilities. As discussed below, this requirement restricts further the allowed Horndeski models \cite{Denissenya:2018mqs}. 

As we have mentioned above, the gravitational properties of Horndeski theories and the corresponding observable quantities are uniquely specified by the four independent $\alpha_i(a)$ functions 
\cite{DeFelice:2011hq,Bellini:2014fua,Sawicki:2015zya}) 
and the background expansion rate $H(a)$. These quantities in turn are determined by the form of the Horndeski Lagrangian density as discussed in the next section and may be used to reconstruct them. The $\alpha_i$ functions are connected not only with the fundamental Horndeski Lagrangian density but also with gravitational observables like the (dimensionless, reduced) gravitational coupling entering the growth of perturbations 
$\mu(a)\equiv G_{\rm growth}(a)/G$ (by $G$ we mean here the usual numerical value of Newton's constant) 
and lensing properties $\Sigma(a)\equiv G_{\rm lensing}(a)/G$
where $G_{\rm growth}$, resp. $G_{\rm lensing}$, is the effective gravitational coupling for the growth of cosmological perturbations, resp. for lensing 
($8\pi G\equiv 1/M_p^2$ where $M_p$ is the (reduced) Planck mass). 
The numerical value of $G$ is obtained from local experiments (solar system, Eotvos type). 
Of course, depending on the models, these gravitational couplings can have a broader physical meaning. For example, in massless scalar-tensor models $G_{\rm growth}$ was called $G_{\rm eff}$, the effective coupling for Newton's gravitational attraction law in a laboratory experiment \cite{Boisseau:2000pr,EspositoFarese:2000ij}. 
An efficient way to explore the physical content of Horndeski models as well as observational constraints on these theories it to parametrize the $\alpha_i$ functions \cite{Espejo:2018hxa,Pace:2019uow}. Such parametrizations usually assume the validity of GR at early times ($\alpha_i(a\simeq 0)=0$) while they allow for a deviation from GR at late times in accordance with the observed accelerating expansion. 
Using such parametrizations, the gravitational strength observables $\mu$ and $\Sigma$ may be derived and compared with cosmological observations leading to constraints on the parameters involved in the evolution of the $\alpha_i$ functions. 
However in view of what was mentioned above, a physically interesting parameter region should satisfy additional requirements beyond consistency with cosmological observations as
it should correspond to viable Horndeski models. 
In this work we investigate stable Horndeski models and we assume an early time behavior consistent with GR, $c_T=c$, scale independence of the $\alpha$ functions on subhorizon scales in the quasi-static approximation (QSA), and finally a background expansion $H(a)$ mimicking $\Lambda$CDM. We assume further a specific dependence of the $\alpha$ functions on $a$, viz. 
$\alpha_i= \alpha_{i0} a^s$, $i=M,B$, where $\alpha_{i0}$ are arbitrary parameters and $s$ is some positive exponent.  

With these assumptions, the goal of the present analysis is to address the following questions:
\begin{itemize}
\item
What is the allowed parameter space for our parametrization of the $\alpha$ functions ?  
\item
Which behavior for $\mu(a)$ and $\Sigma(a)$ is obtained especially at recent times $a \simeq 1$ and is it consistent with observational constraints ? 
\item
How does the growth index $\gamma$ behave in the 
parameter space defining the functions $\alpha_i$ ? 
\end{itemize}
The structure of this paper is the following: In the  next Section \ref{stab} we present a brief review of the Horndeski models. In the context of the $\alpha$ parametrization and the above assumptions, we derive the allowed parameter regions for various values of the exponent $s$. We also obtain the allowed forms of $\mu$ and $\Sigma$,  
 comparing our results with previous studies. 
In Section \ref{constraints} we use  compilations  of  $f\sigma_8$ and $E_G$ data along with  the theoretical expressions for $f\sigma_8$ and $E_G$ statistics data in order to derive constraints on $\mu$ and $\Sigma$ and to obtain  the  allowed range of the functions $\alpha_M(a)$ and $\alpha_B(a)$. 
In Section \ref{flatbackground} 
we consider the growth index $\gamma(z)$ and identify the $(\alpha_{M0},\alpha_{B0},s)$ parameter region that corresponds to specific signs of $\gamma_0-\gamma_0^{\Lambda CDM}$, and $\gamma_1-\gamma_1^{\Lambda CDM}$.
Finally in Section \ref{concl} we conclude, summarize and  discuss  the  implications  of the present analysis.

\section{Stability and generic forms of $\mu$ and $\Sigma$ for viable Horndeski theories}
\label{stab}
 
The Horndeski action, first written down in Ref.\cite{Horndeski:1974wa} and then rediscovered as a generalisation of galileons in Ref. \cite{Deffayet:2011gz,Kobayashi:2011nu}, 
is given by
\be 
S=\int d^4x\sqrt{-g}\left[\sum_{i=2}^5 \mathcal{L}_i\left[g_{\mu\nu},\phi\right]+\mathcal{L}_m\left[g_{\mu\nu},\psi_m\right]\right]
\ee
where the Lagrangian density, $\mathcal{L}_m$, for all matter fields $\psi_m$  is universally coupled to the metric $g_{\mu\nu}$ and does not have direct coupling with the scalar field, $\phi$. The $\mathcal{L}_i$ are the scalar-tensor Lagrangians which depend on the new degree of freedom $\phi$, viz.
\begin{align}
\mathcal{L}_2&=K(\phi,X),\nonumber\\
\mathcal{L}_3&=-G_3(\phi,X)\Box\phi,\nonumber\\
\mathcal{L}_4&=G_4(\phi,X)R+G_{4X}(\phi, X)\Bigl[\left(\Box\phi\right)^2-\nabla_{\mu}\nabla_{\nu}\phi \nabla^{\mu}\nabla^{\nu}\phi\Bigr],\nonumber\\
\mathcal{L}_5&=G_5(\phi,X)G_{\mu \nu}\nabla^{\mu}\nabla^{\nu}\phi-\frac{1}{6} G_{5X}(\phi, X)\Bigl[\left(\Box\phi\right)^3\nonumber\\
&-3\left(\nabla_{\mu}\nabla_{\nu}\phi \right) \left(\nabla^{\mu} \nabla^{\nu}\phi \right)\Box\phi \nonumber\\
&+ 2\left(\nabla_{\mu}\nabla^{\nu}\phi \right)\left(\nabla_{\nu}\nabla^{\beta}\phi \right)\left(\nabla_{\beta} \nabla^{\mu}\phi \right)\Bigr]~,
\end{align}
where $K(\phi,X)\equiv G_2(\phi,X)$ is the K-essence term, $G_i(\phi,X)$ $(i=3,4,5)$ are three coupling functions of the scalar field $\phi$ and its canonical kinetic energy  $X\equiv -\frac{1}{2}\nabla^{\mu}\phi\nabla_{\mu}\phi$, $R$ is the Ricci  scalar, $G_{\mu\nu}$ is the Einstein tensor, $G_{iX}\equiv \partial G_i/\partial X$ and  $G_{i\phi}\equiv \partial G_i/\partial \phi$. In principle the functions $G_i(\phi,X)$ can be chosen freely and determine a particular Horndeski model. 

As mentioned above Horndeski models are characterized by means of four functions of time, $\alpha_i(t)$ $(i=M,K,B,T)$, (see Ref.\cite{Bellini:2014fua}) in addition to the  background evolution encoded in the Hubble parameter $H(a)$. Thus using these functions which fully specify the linear evolution of perturbations 
allows us to disentangle the background expansion from the evolution of the perturbations. The functions $\alpha_K$, $\alpha_B$, $\alpha_T$  are connected to the Lagrangian terms as follows \cite{Bellini:2014fua}
\begin{align}
H^2&M_*^2\alpha_K=2X(K_X+2XK_{XX}-2G_{3\phi}-2XG_{3\phi X})+\nonumber\\
&+12\dot{\phi}XH(G_{3X}+XG_{3XX}-3G_{4\phi X}-2XG_{4\phi XX})+\nonumber\\
&+12XH^2(G_{4X}+8XG_{4XX}+4X^2G_{4XXX})-\nonumber\\
&-12XH^2(G_{5\phi}+5XG_{5\phi X}+2X^2G_{5\phi XX})+\nonumber\\
&+4\dot{\phi}XH^3(3G_{5X}+7XG_{5XX}+2X^2G_{5XXX})
\end{align}
\begin{align}
\label{aB}
HM_*^2\alpha_B&=2\dot{\phi}(XG_{3X}-G_{4\phi}-2XG_{4\phi X})+\nonumber\\
&+8XH(G_{4X}+2XG_{4XX}-G_{5\phi}-XG_{5\phi X})+\nonumber\\
&+2\dot{\phi}XH^2(3G_{5X}+2XG_{5XX})
\end{align}
\begin{align}
M_*^2\alpha_T&=2X(2G_{4X}-2G_{5\phi}-(\ddot{\phi}-\dot{\phi}H)XG_{5X})
\end{align}
Note that we use the definition  $\alpha_B$ of \cite{Bellini:2014fua,Ishak:2019aay}. 
The quantities $\phi$, $X$ and $H$ are evaluated on their background solution to give the particular time-dependence of the $\alpha_i$ functions for that solution. 
Also  $M_*^{-2}$ is proportional to the gravitational coupling entering the cosmological background evolution.
Like in many MG models, it can depend on time and is given by  \cite{Bellini:2014fua} 
\be
M_*^2\equiv2(G_4-2XG_{4X}+XG_{5\phi}-\dot{\phi}HXG_{5X})~,
\label{effmasspl}
\ee
where $\phi$ is the homogeneous value of the scalar field on the cosmic background and a dot denotes differentiation with respect to cosmic time $t$.

Each function $\alpha_i(t)$ is linked with a specific physical property and describes particular classes of models 
In particular, the braiding function $\alpha_B$ describes the mixing of the kinetic terms of the scalar and metric, the kineticity $\alpha_K$ parametrizes the kinetic energy of the scalar perturbations, the tensor speed excess  $\alpha_T$ quantifies how much the gravitational waves (tensor perturbations) speed $c_T$ deviates from that of light, 
finally $\alpha_M$ describes the evolution of $M_*^2$ as follows
\cite{Bellini:2014fua,Gleyzes:2014rba}
\be 
\alpha_M\equiv H^{-1}\frac{d\ln M_*^2}{dt}~.
\label{effmassrate}
\ee
The $\Lambda$CDM model, and more generally GR, corresponds to the particular case 
$M_*^2=M_p^2$ 
and $\alpha_M=\alpha_B=\alpha_K=\alpha_T=0$.

In Horndeski theories, we obtain the Friedmann equations replacing $M_p$ with 
the effective Planck mass 
$M_*$, so the Friedmann equations take the form \cite{Bellini:2014fua,Ishak:2019aay} 
\be 
3H^2=\frac{1}{M_*^2}(\rho_m+\mathcal{E}_{DE})
\label{fried1}
\ee
\be 
2\dot{H}+3H^2=-\frac{1}{M_*^2}(p_m+\mathcal{P}_{DE})
\label{fried2}
\ee
where $\mathcal{E}_{DE}$ and $\mathcal{P}_{DE}$ are the energy density and pressure associated to the additional degree of freedom (the full expressions are provided in the Appendix \ref{Appendix_A}). They are related to the energy density $\rho_{DE}$ and pressure $P_{DE}$ of the effective dark energy component as,
\begin{align}
    \rho_{DE}&=\mathcal{E}_{DE}-3(M_*^2-M_p^2)H^2\\
    P_{DE}&=\frac{M_p^2}{M_*^2}\mathcal{P}_{DE}
\end{align}
where in the last expression we have put $p_m=0$ as we consider here dust-like matter. 
With these definitions the modified Friedmann equations are recast into an Einsteinian form, viz. 
\be 
3H^2=\frac{1}{M_p^2}(\rho_m+\rho_{DE})~,
\label{fried1E}
\ee
\be 
2\dot{H}+3H^2=-\frac{1}{M_p^2} P_{DE}~.
\label{fried2E}
\ee

The stability conditions to be imposed on the functions $\alpha_i(a)$ are the following \cite{Bellini:2014fua,Denissenya:2018mqs}
\be
\alpha_K+\frac{3}{2}\alpha_B^2\geq 0 
\label{akfun} 
\ee
\be
c_s^2>0
\label{cscon}
\ee
where  $c_s$ is the speed of sound which is connected to the $\alpha_i$'s as follows \cite{Bellini:2014fua,Gleyzes:2014rba}
\begin{align}
&\left(\alpha_K+\frac{3}{2}\alpha_B^2\right)c_s^2=
\frac{\dot{\alpha}_B}{H}
-\frac{\rho_m}{H^2 M_*^{2}}
\nonumber\\
&-\left(2-\alpha_B\right)\Bigl[\frac{\dot{H}}{H^2}+\alpha_T-\alpha_M-\frac{\alpha_B}{2\left(1+\alpha_T\right)}\Bigr]~.
\label{cs}
\end{align}
The gravitational waves travel at the speed (with $c=1$)
\be
c_T^2=1+\alpha_T~.
\ee
Recent multimessenger constraints on gravitational waves using the neutron star inspiral GW170817 detected through both the emitted gravitational waves and $\gamma$-rays GRB 170817A  \cite{TheLIGOScientific:2017qsa,Goldstein:2017mmi,Savchenko:2017ffs,Monitor:2017mdv}, imply that $c_T$ is extremely close to the speed of light i.e. $c_T=1\pm 10^{-15}$. This constraint effectively eliminates all Horndeski theories with  $\alpha_{T,0}\equiv \alpha_T(a=1)\neq 0$ (we take $a_0=1$). 
We consider in this paper only those models satisfying $\alpha_T=0$. 

The  $\alpha_i$ functions are independent of each other, i.e.  they can be parametrized independently. However, for simplicity and in accordance with previous studies \cite{Kennedy:2018gtx,Denissenya:2018mqs},
we assume that all the functions $\alpha_i$ 
have the same power law dependence on the scale factor $a$, viz.
\be
\alpha_i=\alpha_{i0}~a^s \quad \textrm{with}  \quad  s>0   
\label{param}
\ee
where the constants $\alpha_{i0}$ are their current values.
The exponent $s$ determines the time evolution for the 
considered modified gravity model.
One of the main goals of this analysis is to impose constraints on these parameters using cosmological observations and the assumptions mentioned earlier.
From \eqref{effmassrate} we have for the quantity $M_*$
\be
M_*=M_p e^{\int_0^a \alpha_M\frac{da'}{2a'}}=M_p e^{\alpha_{M0}\frac{a^s}{2 s}}~,\label{M*}
\ee
in accordance with our assumption $M_*(a=0)=M_p$. 
We obtain also 
\be
M_*(a=1) = M_p e^{\frac{\alpha_{M0}}{2 s}}~.\label{M*0}
\ee 
We have therefore $M_*(a=1)\approx M_p$ for $\alpha_{M0}\ll 2s$. Otherwise, the local value of the scalar field $\phi$ must differ from its value on cosmic scales. We recover 
$\frac{ \dot{M_*} }{ M_* } = \frac{\alpha_M}{2} ~H$ in accordance with \eqref{effmassrate}, and in particular 
\be  
\frac{ \dot{M_*} }{ M_*}(a=1) = \frac{\alpha_{M0}}{2} ~H_0~.\label{dotM*}
\ee  
On subhorizon scales, the QSA applies to  scales  below  the  sound horizon  of  the  scalar  field ($k\gg aH/c_s$ or $\lambda\ll \lambda_J$ where $\lambda_J$ is the Jeans length) \cite{Boisseau:2000pr,Tsujikawa:2007gd,DeFelice:2011hq} and the time-derivatives of the metric and of the scalar field perturbations are neglected compared to their spatial gradients.  
In the conformal Newtonian gauge, the perturbed Friedmann-Lema\^{i}tre-Robertson-Walker (FLRW) 
metric takes the form
\be  
ds^2=-(1+2\Psi)~dt^2+a^2(1-2\Phi)~d \vec{x}^2
\label{metric} 
\ee  
This leads to the following equations for the Bardeen potentials in Fourier space defining our functions $\Sigma(a,k)$ and $\mu(a,k)$
\be 
k^2(\Psi+\Phi)=-8\pi G~\Sigma(a,k)~a^2\rho_m\Delta 
\label{poissonsigma} 
\ee  
\be  
k^2\Psi=-4\pi G~\mu(a,k)~a^2\rho_m\Delta~.
\label{poissonmu} 
\ee
In these equations  $\rho_m$ is the background matter density and $\Delta$ is the comoving matter density contrast defined as $\Delta\equiv\delta_m+3Ha\upsilon/k$ , with $\delta_m\equiv\delta\rho_m/\rho_m$ the matter density contrast in the Newtonian conformal gauge and $\upsilon$ the  irrotational component of the peculiar velocity \cite{Boisseau:2000pr}. 
The functions $\Sigma(a,k)$ and $\mu(a,k)$ are generically time and scale dependent 
encoding the possible modifications of GR defined as\footnote{Note that the precise  definitions of $\Sigma$ and $\mu$  may vary in the literature  (e.g. in Ref. \cite{Ishak:2019aay}).} 

\be  
\mu(a,k)\equiv\frac{G_{\rm growth}(a,k)}{G}  
\ee
\be 
\Sigma(a,k)\equiv \frac{G_{\rm lensing}(a,k)}{G}
\ee
where $G$ is Newton's constant as measured by local experiments, $G_{\rm growth}$ is the effective 
gravitational coupling which is related to the growth of matter perturbation and $G_{lensing}$ is the effective gravitational coupling associated with lensing. 
Anisotropic stress between the gravitational potentials $\Psi$ and $\Phi$ is produced from the Planck mass run rate $\alpha_M$ and the tensor speed excess $\alpha_T$ \cite{Saltas:2014dha}.

Using the gravitational slip parameter $\eta$ (or anisotropic parameter) defined as
\be 
\eta(a,k)=\frac{\Phi(a,k)}{\Psi(a,k)} 
\label{slip}
\ee 
and the ratio of the Poisson equations  (\ref{poissonsigma}), (\ref{poissonmu}),  the two functions  $\mu$ and $\Sigma$ are related as
 \be 
\Sigma(a,k)=\frac{1}{2}\mu(a,k)\left[1+\eta(a,k)\right]
\label{smh} 
\ee
In GR  we have $\mu=1$, $\eta=1$ and $\Sigma=1$. The deviations from GR are expressed by allowing for a scale and time dependent $\mu$ and $\Sigma$ but in  the  present  analysis we ignore scale dependence in the context of the QSA  
and also due to the lack of good quality scale dependent data.

In the case of Horndeski modified gravity, in the quasistatic limit and fixing $\alpha_T=0$ at all times, the functions $\mu(a)$ and $\Sigma(a)$ take the form \cite{Ishak:2019aay}
\be 
\mu(a)=\frac{M_p^2}{M_*^2}\left[1+\frac{2\left(\alpha_M+\frac{1}{2}\alpha_B\right)^2}{c_s^2\left(\alpha_K+\frac{3}{2}\alpha_B^2\right)}\right]
\ee

\be 
\Sigma(a)=\frac{M_p^2}{M_*^2}\left[1+\frac{\left(\alpha_M+\frac{1}{2}\alpha_B\right)\left(\alpha_M+\alpha_B\right)}{c_s^2\left(\alpha_K+\frac{3}{2}\alpha_B^2\right)}\right]
\ee

Thus for theories with $\alpha_M=0$  or $\alpha_B=-2\alpha_M$, $\mu$  is equivalent to $\Sigma$. Notice also that for $\alpha_M=0$, we obtain $\mu>1$ and $\Sigma>1$. The case $\alpha_B=-2\alpha_M$ is a special case also known as No slip Gravity \cite{Linder:2018jil} for which $\eta=1$ and we have then 
\be 
\mu(a)=\Sigma(a) =\frac{M_p^2}{M_*^2}~.
\label{noslip}
\ee
Notice also that all expressions depend on the coefficient $c_s^2\left(\alpha_K+\frac{3}{2}\alpha_B^2\right)$ which from Eq.(\ref{cs}) shows that $\mu$ and $\Sigma$ are actually independent of $\alpha_K$. This parameter has minimal effect on subhorizon scales (i.e. $k/aH\gg 1$) \cite{Bellini:2014fua,Denissenya:2018mqs}, while being uncorrelated with all other functions $\alpha_i$ \cite{Reischke:2018ooh}. It is only independently constrained by stability considerations through Eq. (\ref{akfun}). In addition, as we set $\alpha_T=0$ at all times, the only functions that can be constrained with observations by the quantities $\mu$ and $\Sigma$ are the functions $\alpha_M(a)$ and $\alpha_B(a)$. Finally, assuming the stability conditions (\ref{akfun}), (\ref{cscon}), we have $\mu>M_p
^2/M_*^2$ as noticed in \cite{Amendola:2019laa} but $\Sigma$ remains unconstrained.

For any $w_{DE}(a=\infty)=w_\infty$ finite, we can consider two cases, depending on the sign of $\alpha_{M0}$. In the asymptotic future, the Hubble function evolves as $H\propto a^{-3(1+w_\infty)/2}$ and therefore $\dot H/H^2\rightarrow -3(1+w_\infty)/2$. Also $\dot\alpha_B/H=s~\alpha_B$ where we have assumed Eq.(\ref{param}). It is therefore easy to show that for large scale factor, we have
\begin{align}
    (\alpha_K+\frac{3}{2}\alpha_B^2) c_s^2\rightarrow & s\alpha_B-\frac{3}{2}(1+w_\infty)\alpha_B -\frac{\rho_m}{H^2 M_*^2}\nonumber\\
    & -\alpha_B(\alpha_M+\frac{\alpha_B}{2})
\end{align}
The first two terms are always negligible compared to the last term, except for No Slip Gravity for which the last term is absent.
%
If $\alpha_{M0}<0$, the coefficient $-\rho_m/H^2 M_*^2=-\rho_m e^{-\alpha_{M0}s^s/s}/H^2M_p^2$ is dominant because of the exponential behavior and hence $c_s^2$ is always negative for $a\rightarrow \infty$, 
\begin{align}
    (\alpha_K+\frac{3}{2}\alpha_B^2) c_s^2\propto -\rho_m/H^2 M_*^2<0\,,
    \label{cas1}
\end{align}
and these models are excluded. 
On the other hand if $\alpha_{M0}>0$, the matter component $-\rho_m/H^2 M_*^2=-\rho_m e^{-\alpha_{M0}a^s/s}/H^2M_p^2$ is negligible, we have for $a\rightarrow \infty$
\be
(\alpha_K+\frac{3}{2}\alpha_B^2)c_s^2 \simeq -\frac{a^{2s}}{2}\alpha_{B0}(\alpha_{B0}+2\alpha_{M0})~,
 \label{cas2}
\ee
from which we obtain the condition 
\be
\alpha_{B0}(\alpha_{B0}+2\alpha_{M0})\leq 0~.
\ee
Therefore, we conclude that the only possible viable sector 
satisfies 
\be
\alpha_{B0}\leq 0 ~~~~~{\rm and}~~~~~\alpha_{M0}\geq -\alpha_{B0}/2~.
\ee
Considering these restrictions we have at any time 
\begin{align}
\label{musigma}
    \mu(a)\geq\Sigma(a).
\end{align}
In the case of No Slip Gravity, we have in the asymptotic future
\begin{align}
    (\alpha_K+\frac{3}{2}\alpha_B^2) c_s^2\rightarrow \Bigl[s-\frac{3}{2}(1+w_\infty)\Bigr]\alpha_B -\frac{\rho_m}{H^2 M_*^2}
\end{align}
As previously, $\alpha_{M0}<0$ is excluded because of the matter sector which produces a negative contribution. If $\alpha_{M0}>0$, we need to impose the condition $s-\frac{3}{2}(1+w_\infty)\geq 0$, which is irrelevant only if the asymptotic future is phantom $w_\infty<-1$, or if $w_\infty=-1$ which reduces to $s\geq 0$. 
Notice also that if $\alpha_{B0}<0$ and assuming $c_T=1$, we have from Eq.(\ref{aB}) $2\dot\phi (X G_{3X}-G_{4\phi})<0$. This condition reduces to $dF/dt>0$ for scalar-tensor theories for which $G_4=F(\phi)$ and $G_3=0$.

\section{Reconstruction of the $\alpha_M$, $\alpha_B$ functions from observational constraints on $\mu$, $\Sigma$}
\label{constraints}
In the spirit of this formalism disentangling the background from the  perturbations, our background will be fixed. We assume the most conservative and realistic background, $\Lambda$CDM. Therefore,
observational constraints come only from perturbations. We focus on the linear growth of matter perturbations

\be  {\ddot \delta}_m+2H\dot{\delta}_m-4\pi G ~\mu(a) ~\rho_m \delta_m = 0~.
\label{eq:odedeltat}
\ee
In terms of redshift, Eq. (\ref{eq:odedeltat}) takes the following form 
\cite{Boisseau:2000pr,Gannouji:2006jm,Nesseris:2017vor}

\begin{align}
\delta_m'' + \Bigl[\frac{(H^2)'}{2~H^2}  -
{1\over 1+z}\Bigr]\delta_m'-{3\over 2} \frac{(1+z)~\Omega_{m,0}
~ \mu(z)
}{H^2/H_0^2}~\delta_m=0
\label{eq:odedeltaz}
\end{align}
where a prime denotes differentiation with respect to the redshift.

Note that we have defined $\Omega_m=\frac{\rho_m}{3M_p^2 H^2}$. This definition assumes that general relativity is recovered at small scales. 
Therefore we presume a sufficient viable screening mechanism. It is important to notice that even if we have defined a power law dependence of the parameters (see eq. \ref{param}), the Lagrangian is not totally fixed, principally because of an unconstrained $\alpha_k$. The reconstruction of the Lagrangian from $(\alpha_B, \alpha_M)$ is incomplete and therefore, the Lagrangian is left partially undefined. This freedom can be used to have additional non-linear operators in order to have a viable Vainstein mechanism. Notice that in the static and spherically symmetric case, non-linear operators can be sufficient to eliminate the fifth force and recover general relativity at small scales as shown in \cite{DeFelice:2011th} but in a generic shift-symmetric k-mouflage model, the authors of \cite{Babichev:2011iz} (see also \cite{Kimura:2011dc} for explicit models) have shown that even if the fifth force is suppressed, a time dependence of the scalar field inside the Vainshtein radius remains and therefore at small scales $G_{\text{growth}}=1/8\pi M_*^2(\phi(t))$ 
where $\phi(t)$ is the cosmological time evolution of the scalar field \footnote{In this case, we would have a very strong constraint on the model. Because $|\dot G_{\text{growth}}/G_{\text{growth}}|=|\alpha_M| H$ and considering the Lunar Laser Ranging experiments constrain \cite{Williams:2004qba} $|\dot G/G|<0.02 ~H_0$, we would have $|\alpha_{M0}|<0.02$ because at small scales $G_{\text{growth}}$ should be identified with the gravitational constant $G$. But this result does not apply when the shift symmetry is broken like e.g. in the presence of a mass term.}.
Nevertheless, considering a non spherical problem, general relativity is recovered at small scales \cite{Dar:2018dra}. In conclusion, the screening mechanism could be sufficient to recover general relativity at smaller scales. But it remains a delicate point and should be studied more extensively in the future.

It is usually convenient to introduce the growth function 
\be
f\equiv \frac{{\rm d}\ln \delta_m}{{\rm d}\ln a}~,
\label{growthrate}
\ee
from which it is straightforward to construct the growth index $\gamma$ defined by  
\be
f=\Omega_m^\gamma~. 
\ee
We will constrain the parameters through the growth data $f\sigma_8$ obtained from Redshift Space distortions (RSD) \cite{Macaulay:2013swa,Johnson:2015aaa,Tsujikawa:2015mga,Sola:2016zeg,Wang:2016lxa,Basilakos:2017rgc,Nesseris:2017vor,Kazantzidis:2018rnb,Kazantzidis:2019dvk,Skara:2019usd} and the combination of the growth rate - weak lensing data expressed through the quantity $E_G$ statistics \cite{Joudaki:2017zdt,Amon:2017lia,Leonard:2015cba,Skara:2019usd}.
For a parametrization of $\mu$ and initial conditions deep in the matter era where GR is assumed to hold with $\delta_m \sim a$, equation (\ref{eq:odedeltaz}) may be easily solved numerically leading to a predicted form of $\delta_m(z)$ for a given $\Omega_{m,0}$ and background expansion $H(z)$. 
Once this evolution of $\delta_m$ is known, the observable product 
\be
f\sigma_8(z)\equiv f(z)\cdot \sigma_8(z) = f(z)\cdot \sigma_8 \frac{\delta_m(z)}{\delta_{m,0}}
\ee
can be obtained, 
where $\sigma_8(z)$ is the redshift dependent rms  fluctuations of the linear density field within spheres of (comoving) radius $R=8 h^{-1} Mpc$ while $\sigma_8$ is its value today.
We obtain finally
\be
\fss=-(1+z)\sigma_8 \frac{\delta_m'(z)}{\delta_{m,0}}
\label{eq:fs8}
\ee
This theoretical prediction may now be used to compare with the observed \fs data.

For given parametrizations of our models, we can constrain the function $\Sigma$ (associated to lensing) using $E_G(a)$ data where the observable $E_G(a)$ is defined as 
\cite{Amendola:2012ky,Motta:2013cwa,Pinho:2018unz}
\be 
E_G=\frac{\Omega_{m,0}~\Sigma}{f(z)}
\label{Egth}
\ee
 
This equation assumes that the redshift of the lens galaxies can be approximated by a single value while $E_G$ corresponds to the average value along the line of sight \cite{Pinho:2018unz}. 
Using Eq. (\ref{Egth}) and assuming a specific parametrization for $\alpha_B$ and $\alpha_M$, and a given background expansion, we can compare the theoretical prediction for $E_G$ with the observed $E_G$ datapoints in order to constrain our parameters $(\alpha_{B0},\alpha_{M0})$.
The $f\sigma_8(z)$ and $E_G(z)$ updated data compilations used in our analysis are shown in Tables \ref{tab:data-rsd}  and \ref{tab:data-EG} of the Appendix \ref{sec:Appendix_B} along with the references where each datapoint was originally published. 

We construct $\chi^2_{f\sigma_8}$ and $\chi^2_{E_G}$  as usual \cite{Verde_2010} for the \fs and $E_G$ datasets.  
For the construction of $\chi^2_{f\sigma_8}$ we use the vector \cite{Kazantzidis:2018rnb}
\be 
V_{f\sigma_8}^i(z_i,p)\equiv f\sigma_{8,i}^{obs}-\frac{f\sigma_8^{th}(z_i,p)}{q(z_i,\Omega_{m,0},\Omega_{m,0}^{fid})}
\ee 
where $f\sigma_{8,i}^{obs}$ is the the value of the $i$th datapoint, with $i= 1,...,N_{f\sigma_8}$ ($N_{f\sigma_8}=35$ corresponds to the total number of datapoints of Table \ref{tab:data-rsd}) and $f\sigma_8^{th}(z_i,p)$ is the theoretical prediction, both at  redshift $z_i$. The parameter vector $p$ corresponds to the free parameters $\sigma_8,\Omega_{m,0},\alpha_{B0},\alpha_{M0},s$ that we want to determine from the data.

The fiducial Alcock-Paczynsk correction factor $q$ \cite{Macaulay:2013swa,Nesseris:2017vor,Kazantzidis:2018rnb}  is defined as 
\be  
q(z_i,\Omega_{m,0},\Omega_{m,0}^{fid})=\frac{H(z_i)d_A(z_i)}{H^{fid}(z_i)d_A^{fid}(z_i)}
\label{corr} 
\ee
where  $H(z), d_A(z)$ correspond to the Hubble parameter and the angular diameter distance of the true cosmology and the superscript $^{fid}$ indicates the fiducial cosmology used in each survey to convert angles and redshifts to distances when evaluating the correlation function.  
Thus we obtain $\chi_{f\sigma_8}^2$ as  
\be
\chi_{f\sigma_8}^2(\Omega_{m,0},\alpha_{B0},\alpha_{M0},s,\sigma_8)=V_{f\sigma_8}^iF_{f\sigma_8,ij}V_{f\sigma_8}^j
\label{chif}
\ee
where $F_{f\sigma_8,ij}$ is  the  Fisher  matrix (the  inverse  of  the covariance matrix $C_{f\sigma_8,ij}$ of the data) which is assumed to be diagonal with the exception of the $3\times3$ WiggleZ subspace (see \cite{Kazantzidis:2018rnb} for more details on this compilation).
 
Similarly, for the construction of $\chi_{E_G}^2$, we consider the vector
\be  
V_{E_G}^i(z_i,p)\equiv E_{G,i}^{obs}-E_G^{th}(z_i,p)
\ee 
where $E_{G,i}^{obs}$ is the value of the $i$th datapoint, with $i= 1,...,N_{E_G}$ ($N_{E_G}=8$ corresponds to the total number of datapoints of Table \ref{tab:data-EG}), while $E_G^{th}(z_i,p)$ is the theoretical prediction (Eq. (\ref{Egth})), both at redshift $z_i$. 
%
%
Thus we obtain $\chi_{E_G}^2$ as 
\be
\chi_{E_G}^2 (\Omega_{m,0},\alpha_{B0},\alpha_{M0},s)=V_{E_G}^iF_{E_G,ij}V_{E_G}^j
\label{chieg} 
\ee
where $F_{E_G,ij}$ is  the  Fisher  matrix also assumed to be diagonal.

By minimizing $\chi_{f\sigma_8}^2$ and  $\chi_{E_G}^2$ separately and combined as  $\chi_{tot}^2=\chi_{f\sigma_8}^2+\chi_{E_G}^2$ we obtain the constraints on the parameters  $\alpha_{B0}$ and $\alpha_{M0}$. In this work, we fix  $\Omega_{m,0}=0.315$ and $\sigma_8=0.811$ to the \plcdm parameter  values  favoured  by  Planck  2018 \cite{Aghanim:2018eyx} and other geometric probes \cite{Alam:2016hwk,Scolnic:2017caz}. These values are mainly determined by geometric probes which are independent of the underlying gravitational theory. Specifically, we explore our parameter space ($p$) for  $s=0.5,1,1.5,2,2.5,3$.  

\section{Flat $\Lambda$CDM background}
\label{flatbackground}
In what follows, in agreement with the constraints of most geometric probes \cite{Aghanim:2018eyx,Alam:2016hwk,Scolnic:2017caz}, we assume a background Hubble expansion corresponding to a flat \lcdm cosmology 
with $H(z)$ given by 
\be 
H^2(z) =H_0^2\left[\Omega_{m,0}(1+z)^3 +(1-\Omega_{m,0})\right]~,
\label{eq:hzwcdm} 
\ee
where $\Omega_{m,0}$ is the fractional energy density of dust-like matter today. 

\begin{figure*}
\begin{centering}
\includegraphics[width=0.3\textwidth]{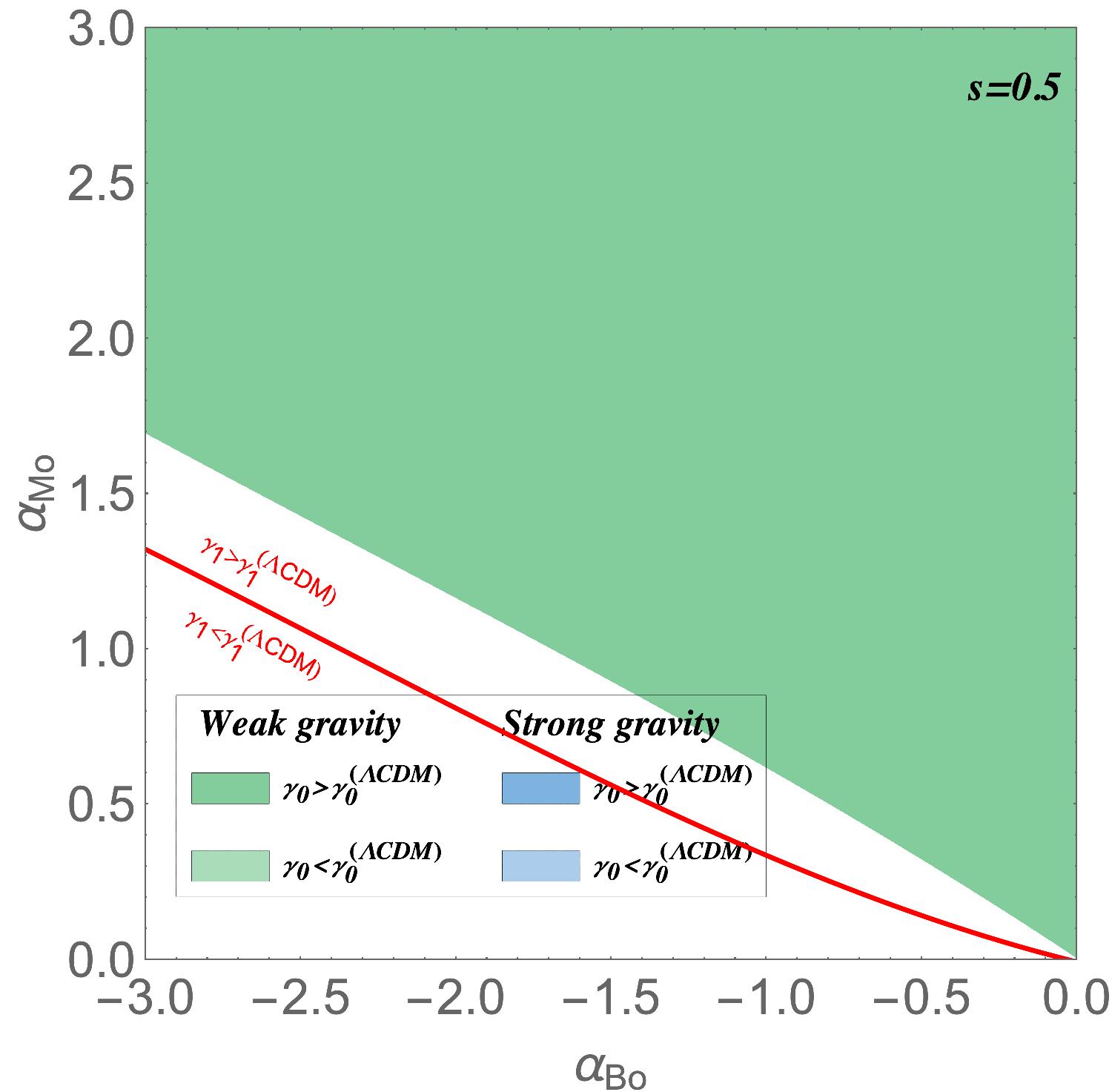}
\includegraphics[width=0.3\textwidth]{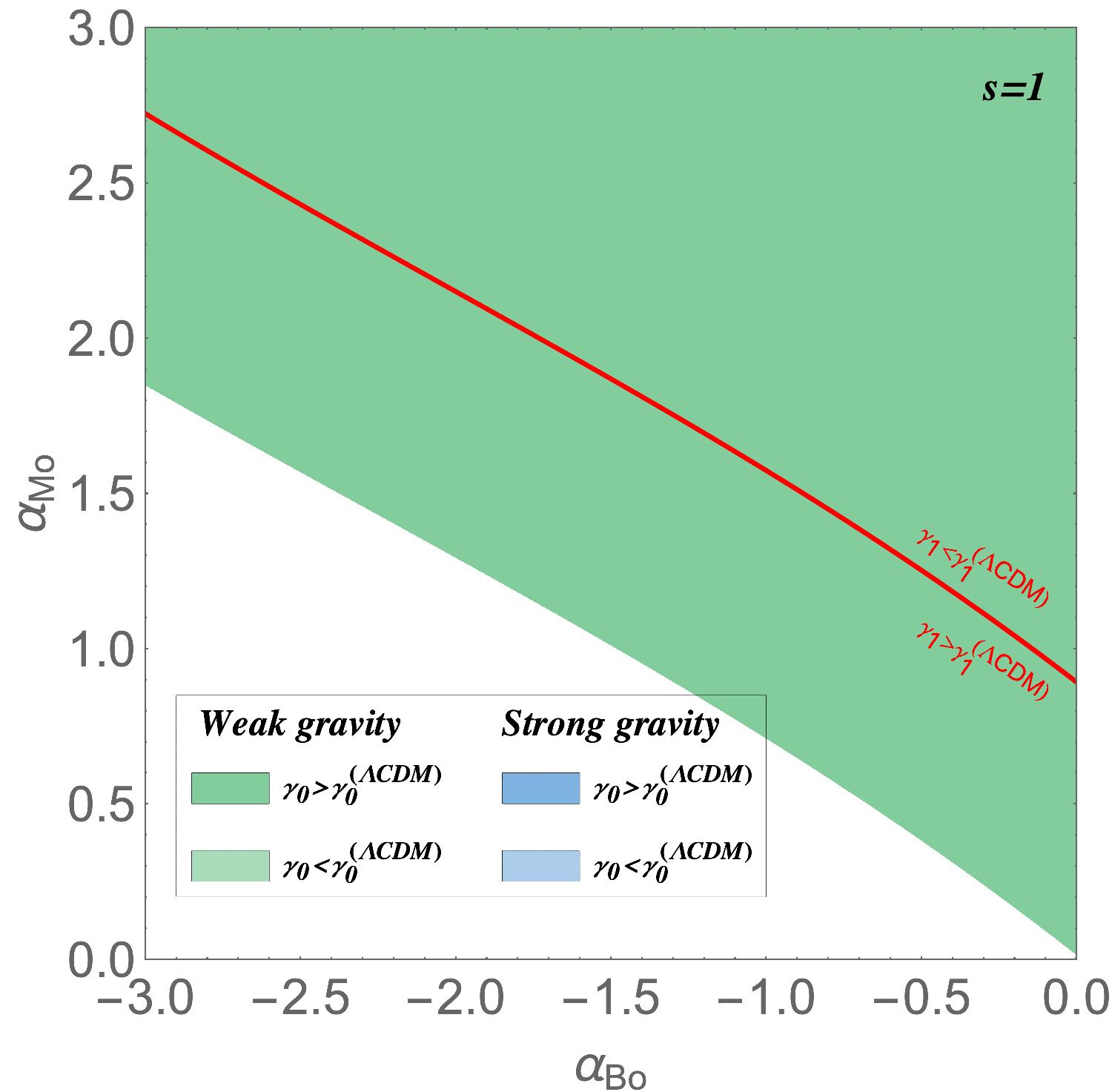}
\includegraphics[width=0.3\textwidth]{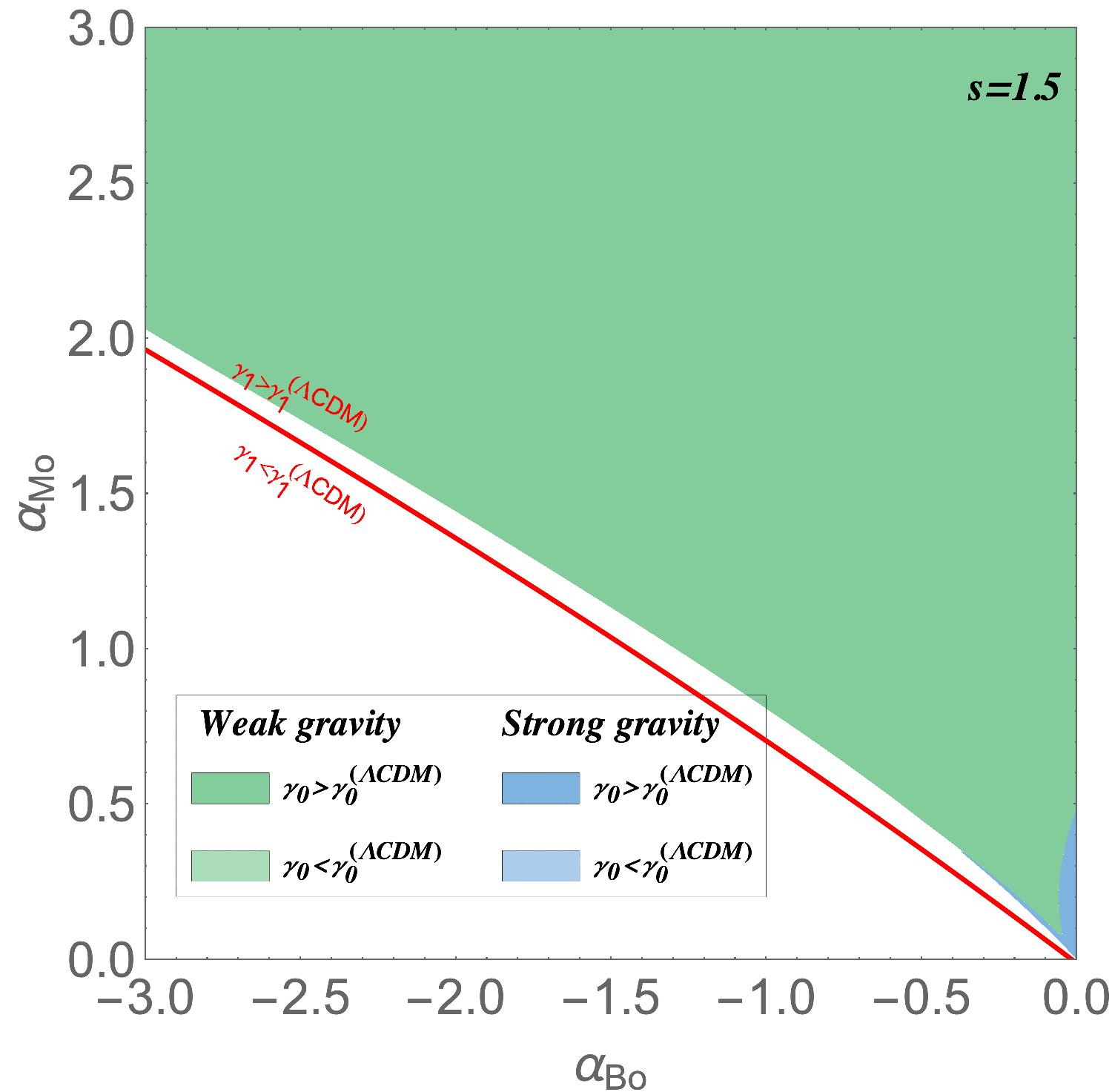}
\includegraphics[width=0.3\textwidth]{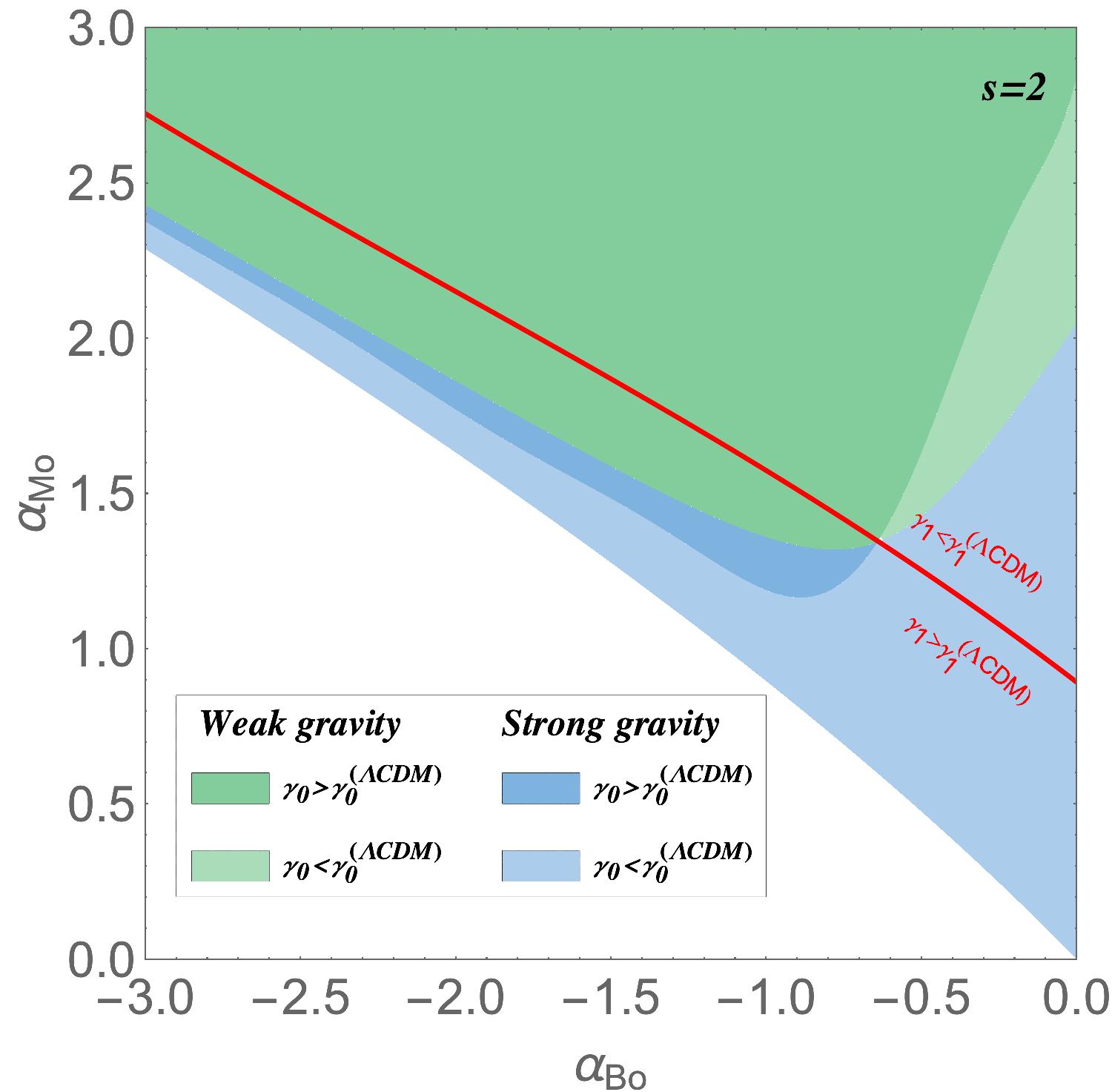}
\includegraphics[width=0.3\textwidth]{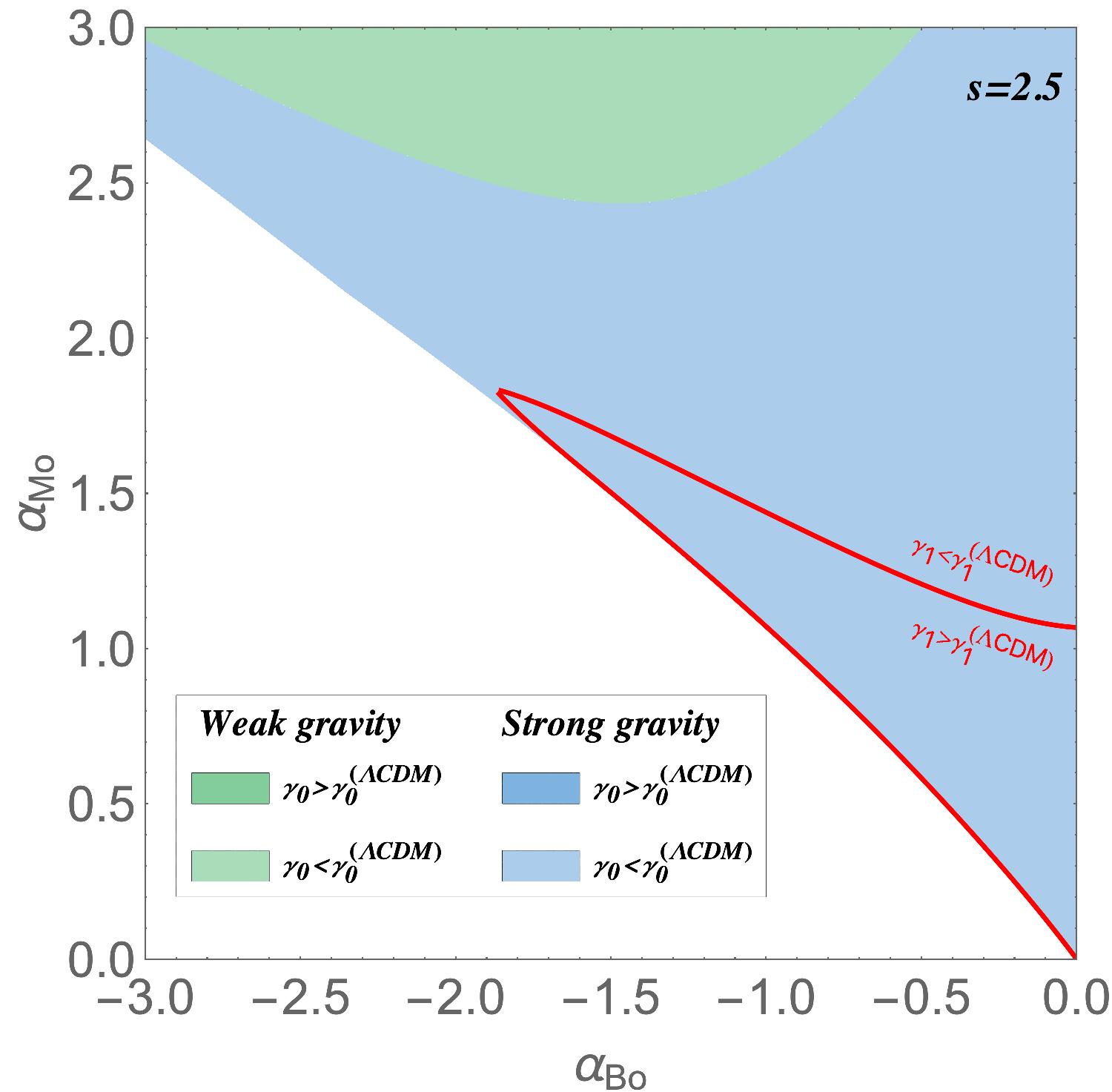}
\includegraphics[width=0.3\textwidth]{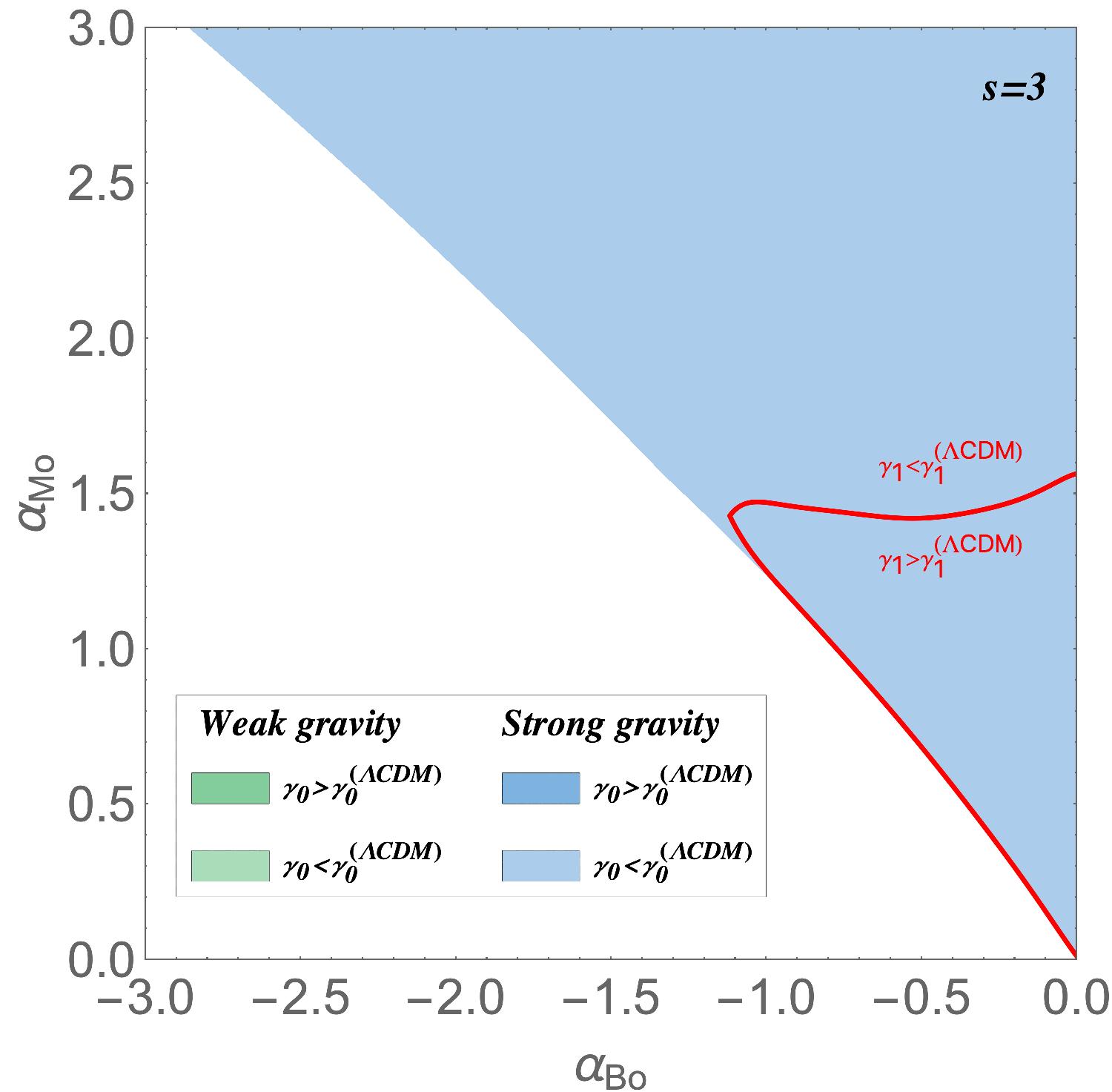}
\par\end{centering}
\caption{The stability (no ghost) region in the 
$\alpha_{M0}$-$\alpha_{B0}$ parameter space is shown and divided into a weak gravity regime today, $\mu_0\equiv \mu(z=0)<1$ (green area), and a strong gravity regime today, $\mu_0>1$ (blue area).
This is obtained by demanding $c_s^2(z)>0$ at all times and assuming a flat $\Lambda$CDM background together with the parametrization Eq. (\ref{param}) used here for the values $s=0.5,1,1.5,2,2.5,3$. 
The dark blue and dark green regions indicate $\gamma_0>\gamma_0^{\Lambda CDM}$, while the light blue and light green regions correspond to $\gamma_0<\gamma_0^{\Lambda CDM}$. 
Finally, the red curve determines the regions where either $\gamma_1>\gamma_1^{\Lambda CDM}$ or $\gamma_1<\gamma_1^{\Lambda CDM}$. 
We see in particular that for $s\leq2$, essentially the weak gravity regime today is selected. In the light green region, $\mu$ crosses $1$ downwards with expansion, while it crosses upwards in the dark blue region.
} 
\label{abamalls}
\end{figure*}

Using the stability equation (\ref{cscon}) (assumed valid for all values of the scale factor $a$) along  with the parametrization  (\ref{param}) for various values of $s$, we show in Fig.(\ref{abamalls}) the stability region (defined by the positivity at all times of the quantity 
$\mu$ and of the sound speed $c_s^2$) in the 
$\alpha_{M0}-\alpha_{B0}$
parameter space. A $\Lambda$CDM background is assumed with a value of $\Omega_{m,0}=0.315$ in accordance with the best fit values of CMB/Planck18\cite{Aghanim:2018eyx}, BAO \cite{Alam:2016hwk} and SNe Ia Pantheon \cite{Scolnic:2017caz} data.

For each region, we show in Fig. \ref{abamalls} the strong gravity regime today
\be
\mu(z=0)>1~,
\ee
and weak gravity regime today 
\be
\mu(z=0)<1~.
\ee
We can see that for small values of ($\alpha_{B0}$,~$\alpha_{M0}$) and
for small $s$, we have $\mu<1$ while for larger $s$ gravity is stronger. Gravity is weak today for $s<2$ and strong if $s>2$ for most of the parameters in the range $-3\leq \alpha_{B0}\leq 0$ and $0\leq \alpha_{M0}\leq 3$.

%
The growth rate of perturbations evolves according to the equation 
\be
\frac{df}{dx} + f^2 + \frac{1}{2} \left(1 - \frac{d \ln \Omega_m}{dx} \right) f = 
                              \frac{3}{2} \frac{G_{growth}}{G} \Omega_m~\label{dfevol}
\ee
where $x\equiv \ln a$. From Eq. (\ref{growthrate}) we have that the density perturbation $\delta_m$ is connected to the growth rate $f$ 
as
\be
\delta_m(a) = \delta_i~{\rm exp} \left[ \int_{x_i}^{x} f(x') dx' \right]~.
\ee
In the special case where the growing mode satisfies $\delta_m\propto a^p$, we have $f=p$ and thus $f\to 1$ in $\Lambda$CDM for large $z$ as long as the decaying mode is negligible \cite{Calderon:2019jem}. In a $\Lambda$CDM universe we have
\be
f=\Omega_m^{\gamma(z)}~,
\ee
with $\gamma_0\equiv \gamma(z=0) \approx \frac{6}{11}$, the latter corresponds to the exact value deep in the matter era and $\gamma_0$ is only slightly higher. In $\Lambda$CDM, $\gamma$ is monotonically increasing with the expansion \cite{Calderon:2019jem}.
In general, the growth index is thus redshift dependent, a strictly constant $\gamma$ being excluded inside GR though it is often 
quasi-constant on redshifts between today till deep in the matter era \cite{Polarski:2016ieb}. 
%
%
Using the above definitions, we have represented in the same figure, the values of the growth index today $\gamma_0$ and its derivative $\gamma_1\equiv \gamma'(z=0)$, where $\gamma_0$, $\gamma_1$ are parameters to be fit to data.

The $\gamma_0,\gamma_1$ values are complementary to the $\mu$ values and add information about the perturbations dynamics in the past. 
On Fig.(\ref{abamalls}), $s=2$, it is seen that the curve $\mu=1$ crosses the curve $\gamma_0=\gamma_0^{\Lambda CDM}$. As we have a fixed $\Lambda$CDM background, it follows from the evolution equation for $\gamma$ that we must have there $\gamma_1=\gamma_1^{\Lambda CDM}$ which is nicely exhibited on our Figure. Furthermore, for that specific point, the value of $\mu$ in the recent past satisfies $\mu\approx 1$ on those redshifts for which $\gamma\approx \gamma_0^{\Lambda CDM} + \gamma_1^{\Lambda CDM}(1-a)$.

Notice that when $\gamma_0=\gamma_0^{\Lambda CDM}$, we are in the weak gravity regime for $\gamma_1<\gamma_1^{\Lambda CDM}$ and in the strong gravity regime for $\gamma_1>\gamma_1^{\Lambda CDM}$. Also if we consider $\gamma_1=\gamma_1^{\Lambda CDM}$, we have $\gamma_0<\gamma_0^{\Lambda CDM}$ for strong gravity and $\gamma_0>\gamma_0^{\Lambda CDM}$ for weak gravity today. These results obtained for our parametrized Horndeski models are in accordance with the results obtained earlier  (see Fig. 7 in \cite{Gannouji:2018col}) in a (gravity) model independent way.  

\begin{figure*}
\begin{centering}
\includegraphics[width=0.24\textwidth]{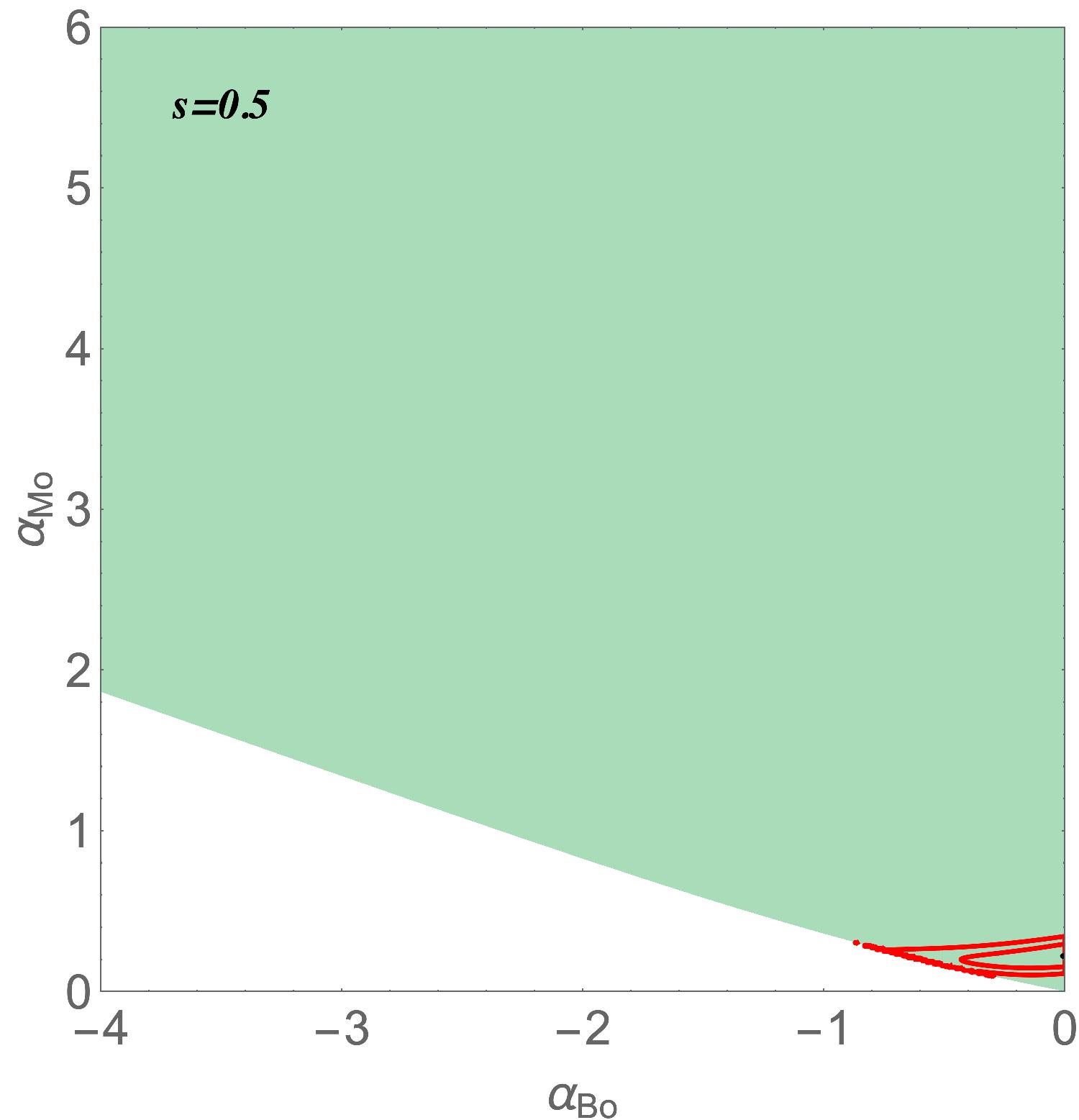}
\includegraphics[width=0.24\textwidth]{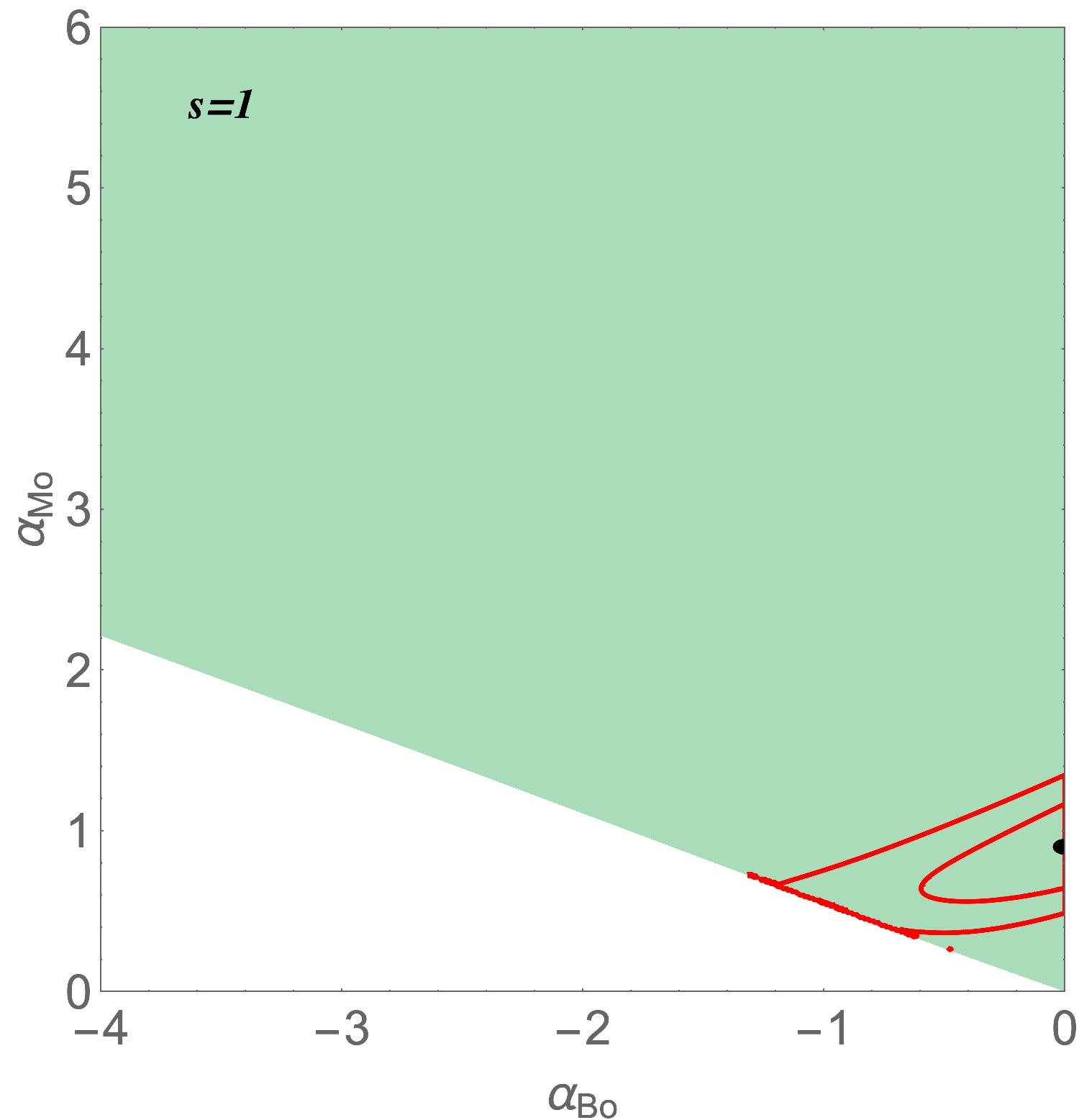}
\includegraphics[width=0.24\textwidth]{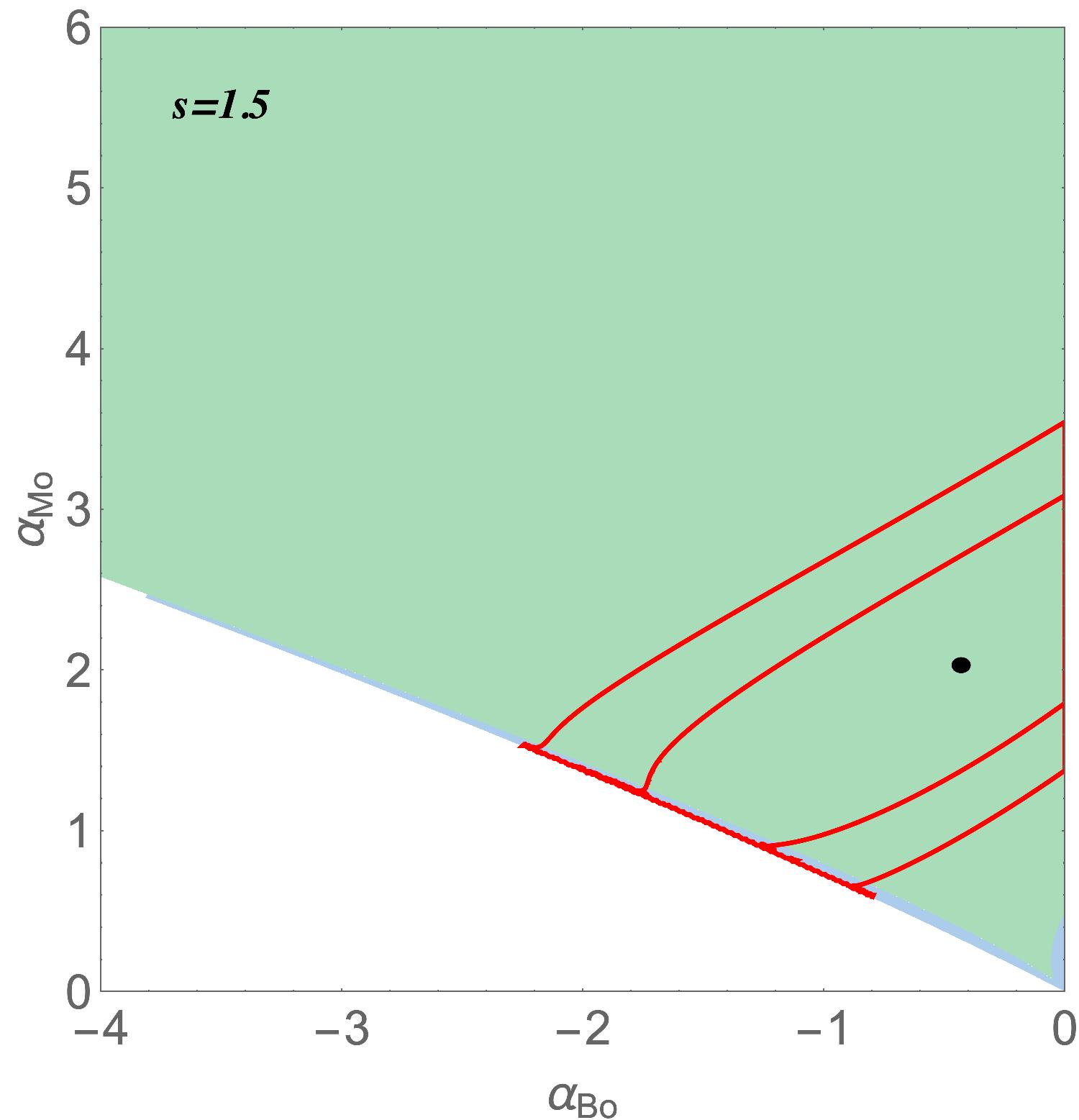}
\includegraphics[width=0.24\textwidth]{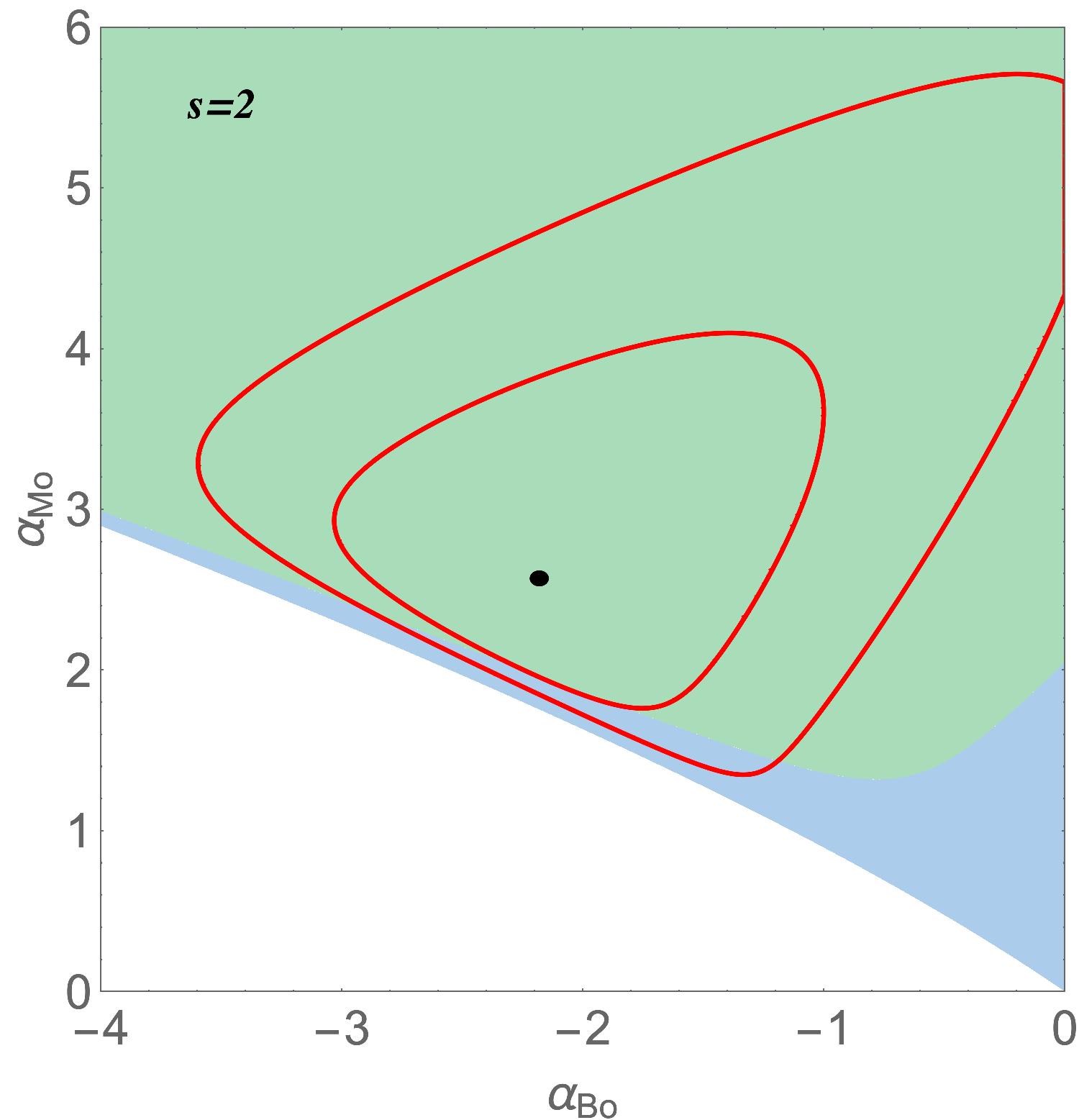}
\par\end{centering}
\caption{The best fit values of $\alpha_{B0}$ and $\alpha_{M0}$ are shown for $s=0.5, 1 ,1.5, 2$ using the combined constraints from the $f\sigma_8$ and $E_G$ data, $1\sigma$ and $2\sigma$ confidence regions are drawned (red curves). As in Fig.\ref{abamalls}, the green area corresponds to weak gravity today while the blue area represents strong gravity today. 
Observations give the constraint $\gamma_0 > \gamma_0^{\Lambda CDM}$ for $s<2$, marginally allowing $\gamma_0 < \gamma_0^{\Lambda CDM}$ for $s=2$.
Note that for higher values of $s$, the best fit moves outside the colored region and is therefore ruled out. 
} 
\label{LCDMconstraints}
\end{figure*}

Using the observational constraints from $f\sigma_8$ and $E_G$ data, we find that for larger values of $s$ ($s\gtrsim 2$) the best fit selects an area violating the stability conditions, and therefore these values of $s$ should be ignored. Therefore, assuming a $\Lambda$CDM background and 
these data, we find that $s\leq 2$ is allowed. This implies that our data select essentially a weak gravity regime today as we have noted earlier, see Fig.(\ref{abamalls}). 
Also, because we have $\Sigma\leq \mu$ from Eq.(\ref{musigma}), we obtain $\Sigma_0\leq 1$, a result which we have confirmed numerically. Even for $s=2$ where a small regime of strong gravity remains, we still have always $\Sigma_0\leq 1$. Fig.(\ref{LCDMconstraints}) exhibits these results with the $1\sigma$ and $2\sigma$ contour plots for the combined data. 

\section{Conclusion - Outlook}
\label{concl} 
Weak gravity is a difficult regime to be reached within viable modified gravity theories. We have shown that assuming a perfectly viable background solution, $\Lambda$CDM, we were able to constrain Horndeski models by using $f\sigma_8$ and $E_G$ data. Assuming only a power law parametrization for the parameters $(\alpha_B,\alpha_M)$, we found that viable models should verify the condition
\begin{align}
    \alpha_{B,0}<0 ~~~~~{\rm and}~~~~~\alpha_{M,0}>-\alpha_{B,0}/2
\end{align}
which constrain $G_{\text{lensing}}$ to be always smaller than $G_{\text{growth}}$.

Considering the $\Lambda$CDM background, we found that for $s<2$, most of the parameters $(\alpha_{B0},\alpha_{M0})$ produce weak gravity today while for $s>2$, we found $\mu_0>1$ for most of the parameters. 
The consideration of cosmological growth data favors $s\leq 2$, namely a mild evolution of the $\alpha_i$ parameters in the late universe, which in turn implies a weak gravity regime today and $\Sigma_0\leq 1$.
We also found that for $s<2$, $\gamma_0>\gamma_0^{\Lambda CDM}$, while for $s>2$ we obtain  $\gamma_0<\gamma_0^{\Lambda CDM}$. Therefore, data also select essentially $\gamma_0>\gamma_0^{\Lambda CDM}$ 
except for $s=2$ for which $\gamma_0<\gamma_0^{\Lambda CDM}$
is marginally allowed.
Note that for $s=2$, in the region with $\mu_0<1$ and $\gamma_0<\gamma_0^{\Lambda CDM}$ (light green on Fig.\ref{abamalls}), gravity was strong in the near past ($z\lesssim 1)$, while in the region 
$\mu_0>1$ and $\gamma_0>\gamma_0^{\Lambda CDM}$ (dark blue on Fig.\ref{abamalls}), gravity was weak in the recent past. 
In some sense, the 
value of $\gamma_0$ indicates that gravity was either weak ($\gamma_0>\gamma_0^{\Lambda CDM}$ and $\mu_0>1$) or strong 
($\gamma_0<\gamma_0^{\Lambda CDM}$ and $\mu_0<1$) when 
we average over the recent past, while $\mu_0$ determines the strength of gravity today. For example, in the light green region where $\mu_0<1$, the average of $\mu(z)$ over redshift $(0,1)$ is larger than 1. We encountered the same behavior in the dark blue region where gravity is strong today and weak on average for most part of the region. Therefore, the pairs $(\gamma_0,\gamma_1)$ add information on the past dynamics of $\mu$.

For models with $\gamma_1=\gamma_1^{\Lambda CDM}$, we have $\gamma_0<\gamma_0^{\Lambda CDM}$ when gravity is strong today and $\gamma_0>\gamma_0^{\Lambda CDM}$ when gravity is weak today. Also when $\gamma_0=\gamma_0^{\Lambda CDM}$, 
we have weak gravity when $\gamma_1>\gamma_1^{\Lambda CDM}$ while we have strong gravity when $\gamma_1<\gamma_1^{\Lambda CDM}$.


In summary, we have proved that under mild assumptions, we could have a consistent and viable weak gravity regime today. It is thus interesting to know how generic this result is. 
It is interesting that the model we have assumed is observationally incompatible at more than $2 \sigma$ with $M_*=M_p$ today (at least for $s \gtrsim 0.5$). Hence the local value of $M_*$ must be necessarily different from its assumed value on cosmic scales today, eq.\eqref{M*0}, and some screening mechanism must be at work in order to make the model viable. As we have mentioned earlier, this is a delicate issue. Even in the absence of screening, our results leave open the possibility to have viable models  with $s<0.5$ satisfying $\alpha_{M0}\ll 2s$. Of course, in that case, $M_*$ would be (very) weakly varying at all times.

We plan to investigate in a future work the relevance, regarding the obtained results, of the two main assumptions made in this work, namely the power-law parametrization of the free functions $\alpha_i$ and the $\Lambda$CDM background expansion (for an alternative approach see e.g. \cite{Garcia-Quintero:2020bac}).  
For example, in the case of minimal scalar-tensor theories, it has been shown that values of $w<-1$ can indeed lead to $\mu<1$ \cite{Gannouji:2018ncm,Kazantzidis:2019dvk}, while it is otherwise impossible to realize. Despite strong restrictions on Horndeski models coming from the gravitational waves speed, viable models could still provide interesting cosmological scenarios with varying gravitational couplings. 

 
\section*{Acknowledgements}
This research is co-financed by Greece and the European Union (European Social Fund - ESF) through
the Operational Programme ”Human Resources Development, Education and Lifelong Learning 2014-2020”
in the context of the project ”Scalar fields in Curved Spacetimes: Soliton Solutions, Observational Results and Gravitational Waves” (MIS 5047648). This  article has also benefited from COST Action CA15117 (CANTATA), supported by COST (European Cooperation in Science and Technology). R.G. is supported by FONDECYT project No 1171384.\\

\appendix
\section{Definitions} 
\label{Appendix_A}
The background 
quantities $\mathcal{E}_{DE}$ and 
$\mathcal{P}_{DE}$
are defined as \cite{Bellini:2014fua}
\begin{widetext} 
\be 
\begin{split}
\mathcal{E}_{DE}\equiv&-K+2X\left(K_X-G_{3\phi}\right)+6\dot{\phi}H\left(XG_{3X}-G_{4\phi}-2XG_{4\phi X}\right)+\\
&+12H^2X\left(G_{4X}+2XG_{4XX}-G_{5\phi}-XG_{5\phi X}\right)+4\dot{\phi}H^3X\left(G_{5X}+XG_{5XX}\right)
\end{split}
\ee
\be 
\begin{split}
\mathcal{P}_{DE}\equiv&K-2X\left(G_{3\phi}-2G_{4\phi\phi}\right)+4\dot{\phi}H\left(G_{4\phi}-2XG_{4\phi X}+XG_{5\phi\phi}\right)-\\
&-M_*^2\alpha_BH\frac{\ddot{\phi}}{\dot{\phi}}-4H^2X^2G_{5\phi X}+2\dot{\phi}H^3XG_{5X}
\end{split}
\ee

\end{widetext}
Note that in the literature there appear various definitions of the energy density associated to the  dark energy (see \cite{Amendola:2019laa}).
However we use here the effective DE energy density $\rho_{DE}$ and pressure $P_{DE}$ based on an Einsteinian representation of modified gravity \cite{Boisseau:2000pr,Gannouji:2006jm}.  

\section{DATA USED IN THE ANALYSIS} 
\label{sec:Appendix_B}

In this appendix we present the data used in the analysis. \\

\begin{widetext} 

\begin{longtable}{ | c | c | c | c | c | c | c | }
\caption{The \fs updated data compilation of Ref. \cite{Skara:2019usd} used in the present analysis.} 
\label{tab:data-rsd}\\
\hline
    Index & Dataset & $z$ & $f\sigma_8(z)$ & Refs. & Year & Fiducial Cosmology \\ 
\hline
\hline

1&2MRS &0.02& $0.314 \pm 0.048$ &  \cite{Davis:2010sw}, \cite{Hudson:2012gt}& 13 November 2010 & $(\Omega_{m},\Omega_K,\sigma_8)=(0.266,0,0.65)$ \\

2 & SDSS-LRG-200 & $0.25$ & $0.3512\pm 0.0583$ & \cite{Samushia:2011cs} & 9 December 2011 & $(\Omega_{m},\Omega_K,\sigma_8)=(0.276,0,0.8)$  \\

3 & WiggleZ & $0.44$ & $0.413\pm 0.080$ & \cite{Blake:2012pj} & 12 June 2012  & $(\Omega_{m},h,\sigma_8)=(0.27,0.71,0.8)$ \\

4 & WiggleZ & $0.60$ & $0.390\pm 0.063$ & \cite{Blake:2012pj} & 12 June 2012 &  \\

5 & WiggleZ & $0.73$ & $0.437\pm 0.072$ & \cite{Blake:2012pj} & 12 June 2012 &\\

6 & GAMA & $0.18$ & $0.360\pm 0.090$ & \cite{Blake:2013nif}  & 22 September 2013 & $(\Omega_{m},\Omega_K,\sigma_8)=(0.27,0,0.8)$ \\

7 & SDSS-MGS & $0.15$ & $0.490\pm0.145$ & \cite{Howlett:2014opa} & 30 January 2015 & $(\Omega_{m},h,\sigma_8)=(0.31,0.67,0.83)$ \\

8 & SDSS-veloc & $0.10$ & $0.370\pm 0.130$ & \cite{Feix:2015dla}  & 16 June 2015 & $(\Omega_{m},\Omega_K,\sigma_8$)$=(0.3,0,0.89)$\cite{Tegmark:2003uf} \\

9 & FastSound& $1.40$ & $0.482\pm 0.116$ & \cite{Okumura:2015lvp}  & 25 November 2015 & $(\Omega_{m},\Omega_K,\sigma_8$)$=(0.27,0,0.82)$\cite{Hinshaw:2012aka} \\

10 & BOSS DR12 & $0.38$ & $0.497\pm 0.045$ & \cite{Alam:2016hwk} & 11 July 2016 & $(\Omega_{m},\Omega_K,\sigma_8)=(0.31,0,0.8)$ \\

11 & BOSS DR12 & $0.51$ & $0.458\pm 0.038$ & \cite{Alam:2016hwk} & 11 July 2016 & \\

12 & BOSS DR12 & $0.61$ & $0.436\pm 0.034$ & \cite{Alam:2016hwk} & 11 July 2016 & \\

13 &VIPERS v7 & $1.05$ & $0.280\pm 0.080$ & \cite{Wilson:2016ggz} & 26 October 2016 &\\

14 &  BOSS LOWZ & $0.32$ & $0.427\pm 0.056$ & \cite{Gil-Marin:2016wya} & 26 October 2016 & $(\Omega_{m},\Omega_K,\sigma_8)=(0.31,0,0.8475)$\\

15 & VIPERS  & $0.727$ & $0.296 \pm 0.0765$ & \cite{Hawken:2016qcy} &  21 November 2016 & $(\Omega_{m},\Omega_K,\sigma_8)=(0.31,0,0.7)$\\

16 & 6dFGS+SnIa & $0.02$ & $0.428\pm 0.0465$ & \cite{Huterer:2016uyq} & 29 November 2016 & $(\Omega_{m},h,\sigma_8)=(0.3,0.683,0.8)$ \\

17 & 2MTF & 0.001 & $0.505 \pm 0.085$ &  \cite{Howlett:2017asq} & 16 June 2017 & $(\Omega_{m},\sigma_8)=(0.3121,0.815)$\\

18 & BOSS DR12 & $0.31$ & $0.384 \pm 0.083$ &  \cite{Wang:2017wia} & 15 September 2017 & $(\Omega_{m},h,\sigma_8)=(0.307,0.6777,0.8288)$\\

19 & BOSS DR12 & $0.36$ & $0.409 \pm 0.098$ &  \cite{Wang:2017wia} & 15 September 2017 & \\

20 & BOSS DR12 & $0.40$ & $0.461 \pm 0.086$ &  \cite{Wang:2017wia} & 15 September 2017 & \\

21 & BOSS DR12 & $0.44$ & $0.426 \pm 0.062$ &  \cite{Wang:2017wia} & 15 September 2017 & \\

22 & BOSS DR12 & $0.48$ & $0.458 \pm 0.063$ &  \cite{Wang:2017wia} & 15 September 2017 & \\

23 & BOSS DR12 & $0.52$ & $0.483 \pm 0.075$ &  \cite{Wang:2017wia} & 15 September 2017 & \\

24 & BOSS DR12 & $0.56$ & $0.472 \pm 0.063$ &  \cite{Wang:2017wia} & 15 September 2017 & \\

25 & BOSS DR12 & $0.59$ & $0.452 \pm 0.061$ &  \cite{Wang:2017wia} & 15 September 2017 & \\

26 & BOSS DR12 & $0.64$ & $0.379 \pm 0.054$ &  \cite{Wang:2017wia} & 15 September 2017 & \\

27 & SDSS-IV & $0.978$ & $0.379 \pm 0.176$ &  \cite{Zhao:2018jxv} & 9 January 2018 &$(\Omega_{m},\sigma_8)=(0.31,0.8)$\\

28 & SDSS-IV & $1.23$ & $0.385 \pm 0.099$ &  \cite{Zhao:2018jxv} & 9 January 2018 & \\

29 & SDSS-IV & $1.526$ & $0.342 \pm 0.070$ &  \cite{Zhao:2018jxv} & 9 January 2018 & \\

30 & SDSS-IV & $1.944$ & $0.364 \pm 0.106$ &  \cite{Zhao:2018jxv} & 9 January 2018 & \\

31 & VIPERS PDR2 & $0.60$ & $0.49 \pm 0.12$ &  \cite{Mohammad:2018mdy} & 6 June 2018 &$(\Omega_b,\Omega_m,h,\sigma_8)=(0.045,0.31,0.7,0.8)$\\

32 & VIPERS PDR2 & $0.86$ & $0.46 \pm 0.09$ &  \cite{Mohammad:2018mdy} & 6 June 2018 & \\

33 & BOSS DR12 voids & $0.57$ & $0.501 \pm 0.051$ &\cite{Nadathur:2019mct} & 1 April 2019 &$(\Omega_b,\Omega_m,h,\sigma_8)=(0.0482,0.307,0.6777,0.8228)$\\

34 & 2MTF 6dFGSv & $0.03$ & $0.404 \pm 0.0815$ &\cite{Qin:2019axr} & 7 June 2019 &$(\Omega_b,\Omega_m,h,\sigma_8)=(0.0491,0.3121,0.6571,0.815)$\\

35 & SDSS-IV & $0.72$ & $0.454 \pm 0.139$ &  \cite{Icaza-Lizaola:2019zgk} & 17 September 2019 &$(\Omega_{m},\Omega_b h^2,\sigma_8)=(0.31,0.022,0.8)$ \\

\hline
\end{longtable}

\begin{longtable}{ | c| c|c|c| c | c | c |c| }
\caption{The $E_G(z)$ data compilation of Ref. \cite{Skara:2019usd} used in the present analysis. }
\label{tab:data-EG}\\
\hline 
  
   Index &Dataset &$z$ & $E_G(z)$  & $\sigma_{E_G}$&Scale  [Mpc/h]  & Reference\\
    
\hline   
\hline  
 
1 &KiDS GAMA&0.267 &0.43& 0.13 & $5<R<40$& \cite{Amon:2017lia}\\

2&KiDS 2dFLenS BOSS LOWZ 2dFLOZ &0.305 &0.27 &0.08 & $5<R<60$& \cite{Amon:2017lia}\\
3&RCSLenS CFHTLenS WiggleZ BOSS WGZLoZ LOWZ&0.32& 0.40 &0.09 & $R>3 $&\cite{Blake:2015vea}\\

4&KiDS 2dFLenS BOSS CMASS 2dFHIZ&0.554 &0.26& 0.07&  $5 < R < 60$ &\cite{Amon:2017lia}\\
5&RCSLenS CFHTLenS WiggleZ BOSS WGZHiZ CMASS&0.57 &0.31& 0.06&  $R>3$& \cite{Blake:2015vea}\\
6&RCSLenS CFHTLenS WiggleZ BOSS WGZHiZ CMASS&0.57 &0.30& 0.07& $ R>10$ &\cite{Blake:2015vea}\\

7&CFHTLenS VIPERS &0.60 &0.16& 0.09& $ 3 < R < 20$ &\cite{delaTorre:2016rxm}\\
8&CFHTLenS VIPERS &0.86 &0.09& 0.07 & $3 < R < 20 $&\cite{delaTorre:2016rxm}\\

\hline
\end{longtable}

\end{widetext}

\raggedleft
\bibliography{bibliography}

\end{document}